\documentclass[amsmath,amssymb,aps,preprint]{revtex4-1}

\usepackage[english]{babel}
\usepackage{graphicx}

\begin{document}
\selectlanguage{english}

\title{$^7$Be and $^{22}$Na radionuclides for a new therapy of cancer}

\author{V.~I.~Kukulin}
\affiliation{Skobeltsyn Institute of Nuclear Physics, Lomonosov
Moscow State University, Leninskie gory, Moscow, Ru-119991,
Russia}
\author{A.~V.~Bibikov}
\affiliation{Skobeltsyn Institute of Nuclear Physics, Lomonosov
Moscow State University, Leninskie gory, Moscow, Ru-119991,
Russia}
\author{E.~V.~Tkalya}
\affiliation{P.N. Lebedev Physical Institute of the Russian
Academy of Sciences, 119991, 53 Leninskiy pr., Moscow, Russia}
\affiliation{National Research Nuclear University MEPhI, 115409, Kashirskoe shosse 31, Moscow, Russia}
\affiliation{Nuclear Safety Institute of RAS, Bol'shaya Tulskaya
  52, Moscow 115191, Russia}
\author{M.~Ceccarelli}
%\email{matteo.ceccarelli@dsf.unica.it}
\affiliation{Department of Physics, University of Cagliari,
  S.P. Monserrato-Sestu km 0.700, I-09042 Monserrato (CA), Italy}

\author{I.~ V.~ Bodrenko}
\email[communications may be sent to ]{igor.bodrenko@dsf.unica.it}
\affiliation{Department of Physics, University of Cagliari,
  S.P. Monserrato-Sestu km 0.700, I-09042 Monserrato (CA), Italy}

%%%%%%%%%%%%%%%%% END OF PREAMBLE %%%%%%%%%%%%%%%%
\begin{abstract} 
  %We discuss the physical grounds and the perspective of using the $^7$Be isotope
  %in the new kind of the neutron-activated radiation therapy of cancer.
  %  As compared to the $^{10}$B nuclide, almost exclusively used in
  The $^{10}$B isotope has been almost exclusively used in
  the  neutron-capture radiation therapy (NCT) of cancer for decades.
  We have identified two other nuclides suitable for the radiotherapy, which have
  ca.10 times larger cross section of absorption for neutrons and emit  heavy charged particles.
  This would provide several key advantages for potential NCT,
  such as the possibility to use either a lower nuclide concentration
  in the target tissues, or a lower neutron irradiation flux. 
  By detecting the characteristic $\gamma$ radiation from the spontaneous decay
  of the radionuclides, one can image and control their accumulation.
  These advantages could be critical for the revival of the NCT as a safer,
  more efficient and more widely used cancer therapy.
\end{abstract}

\maketitle

%{\bf \large One Sentence Summary:} We discuss the physical grounds, the advantages
%and the perspective of using of the two light radionuclides in a new type of teranostics.

\section*{Introduction}

The use of the ionizing radiation in treatment of cancer has a long and successful
history \cite{Citrin2017}, and it is one of the most remarkable examples of direct medical
outreach as results of modern physics.
Energetic photons, leptons or heavy charged particles can ionize biological matter creating  lethal
damages to living cells through the direct disintegration of important biomolecules (like DNA or RNA)
and by creating chemically active radicals destroying cell's biochemistry \cite{Baskar2012}. The radiation
therapy is a valid alternative and a complementary treatment to the surgery and to the chemotherapy in
oncology practices, used in more than 50\% of patients with cancer \cite{Ruysscher2019}.

One of the issues in the radiation therapy is to deliver the ionization selectively into a spatially localized
region occupied by the tumor, ranging from few millimeters to several centimeters in size. The surrounding
healthy tissues must receive as small radiation damage as possible to avoid side effects \cite{Ruysscher2019}.
The ionizing radiation can be delivered from outside the patient, -- a method known as external-beam radiation
therapy (EBRT).
Heavy charged particles (protons, accelerated nuclei) have an advantage over leptons and photons
due to relatively small range straggling and to the sharp Bragg peak in the energy
loss vs the path length curve, located in the end of particle's range in  the tissue
\cite{ziegler1985stopping,Mitin2014,Newhauser2015}.
Therefore, the radiation damage to the tissues may be localized inside the tumor. The heavy particles, however,
have shorter ranges in matter than light particles and photons at the same kinetic energy. Therefore,
to access tumors deep in the body one needs to accelerate the particles up to the energies
of few hundreds MeV/u, to deliver and to focus the beam onto the target in human body. 
This makes the infrastructure for the heavy particle radiation therapy very complex, expensive, and limited
to large medical/research centers, while a much cheaper gamma-ray radiation therapy has come to many
common hospitals \cite{Teasava2009,Mitin2014}.

A way to overcome the problem of delivery of fast particles into tumor is to put the radiation
source directly into the cancerous tissue.
This idea is realized by physically placing a radioactive nuclide enclosed in a protective capsule or as a
wire inside (or next to) the target tissues (brachytherapy), what often requires surgery \cite{Williamson2006}. 
Alternatively, one may use radioactive nuclides bound into radiopharmaceuticals
selectively transported into the target tissues \cite{Jadvar2017}.
The spontaneous decay of a radioisotope nucleus may release up to several
tens of MeV, converted into the kinetic energy of charged product particles
and deposited into the ionization in the range within less than a millimeter around the decaying nucleus. 
The radionuclide therapy, therefore,  replaces the problem of kinematic delivery of the ionizing particle into
the target tissue in EBRT with the problem of selective accumulation of radiopharmaceuticals .
The radionuclide therapy also brings a new issue -- the problematic temporal control. The timescales of the
accumulation and of the excretion of radiopharmaceuticals
(which are regulated by the metabolism and are difficult to control)
should be in due relation to the lifetime of the isotope, so that to maximize the part of the nuclei
decayed in the tumor and not on their way into or out of the target.
It is not possible to initiate and to stop the therapy on an arbitrary moment.
Moreover, many radionuclides as well as their products are heavy elements,
which are known for their toxicity and problematic excretion \cite{Ruysscher2019}.

In the neutron capture radiation therapy (NCT) the active isotope undergoes induced radioactive
decay following the capture of a neutron. This method combines the local energy deposition property of 
the radionuclide therapy and a good temporal control of the beam-particle radiation therapy,
as the neutron flux may be switched on and off quickly.
The idea was suggested in 1936 \cite{locher1936biological} and then implemented for the first time
in 1954 \cite{farr1954neutron}, by utilizing $^{10}$B isotopes and the reaction $^{10}$B(n,$\alpha\gamma$)$^7$Li.
But despite more than 60 years of research and
development, the boron neutron capture therapy (BNCT) is still in the experimental phase
\cite{Barth2005,Barth2012,Slatkin2017,Barth2018}.
Several issues are cited as the reason of this situation, \cite{Moss2014}. First, it is often difficult to identify the
boron-containing compounds that can be selectively accumulated and kept at the necessary
concentration in the tumor cells.
%Second, the thermal neutrons, necessary for efficient BNCT, have small penetration capability, so that
%only tumors close to the surface can be treated. Epithermal neutrons (with kinetic energy, $E_{\mathrm n}>0.5$~eV),
%can penetrate several cm into the body but have smaller absorption cross section on $^{10}$B.
Second, besides many successful cases of practical applications,
a significant number of side effects due to the neutron irradiation of healthy tissues was observed.
Third, the BNCT requires high enough neutron flux ($10^{10}$ to $10^{12}$ cm$^{-2}$s$^{-1}$) available
only on nuclear reactors and on large-scale accelerator complexes until recently, so that the infrastructure is
complex, expensive and not well suited for systematic massive clinical studies. 

Only one other element, gadolinium, mainly $^{157}$Gd isotope in the reaction $^{157}$Gd(n,$\gamma$)$^{158}$Gd,
has been considered for NCT, though in much lesser extent \cite{Cerullo2009}. Indeed the effect of
its huge thermal neutrons absorption cross section (by more than 60 times larger than that for $^{10}$B)
is diminished by the fact that in the most cases the $^{158}$Gd$^*$ excitation energy is taken away from the targeted
tissues by high-energy $\gamma$ photons, which does not create any local ionization. 
On average, only a small part of the excitation energy (less than 1\%) is radiated in the form
of the electrons
via the internal conversion and Auger mechanisms, which is converted into the ionization
within 0.1 mm range from the source \cite{Enger2013}. 
As a result, GdNCT has no obvious benefits over BNCT and still remains in the experimental phase,
mostly focused on the identification of appropriate tumor-selective Gd delivery agents or
of hybrid B-Gd-containing compounds \cite{Deagostino2016}. 

In the present work, we have identified two new nuclides, $^7$Be and $^{22}$Na, suitable for the radiation therapy, 
and suggest to consider them for potential use in the novel NCT.

\section*{Nuclides properties}

\begin{figure}[htb] 
  \begin{center}
    \begin{tabular}{cc}
      \includegraphics[width=0.4\textwidth]{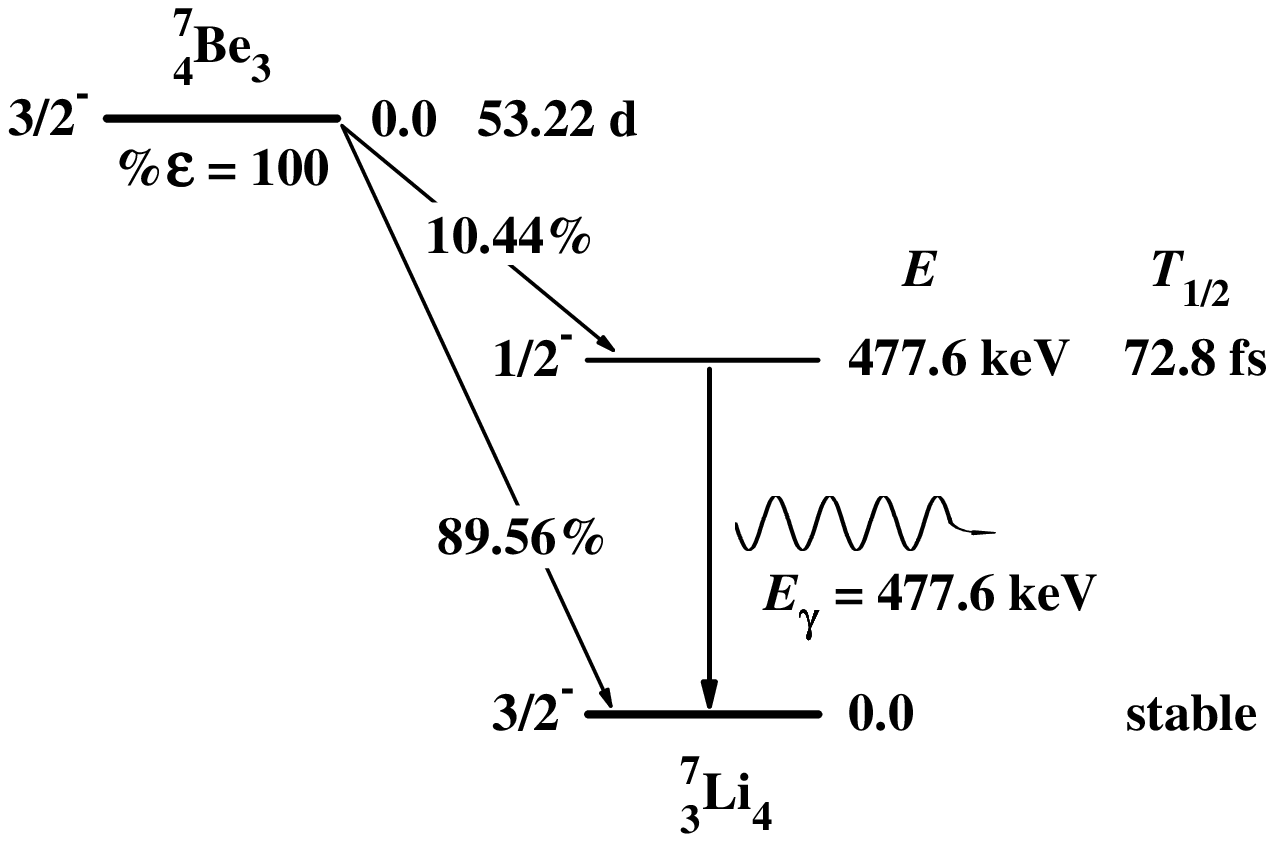} & \includegraphics[width=0.4\textwidth]{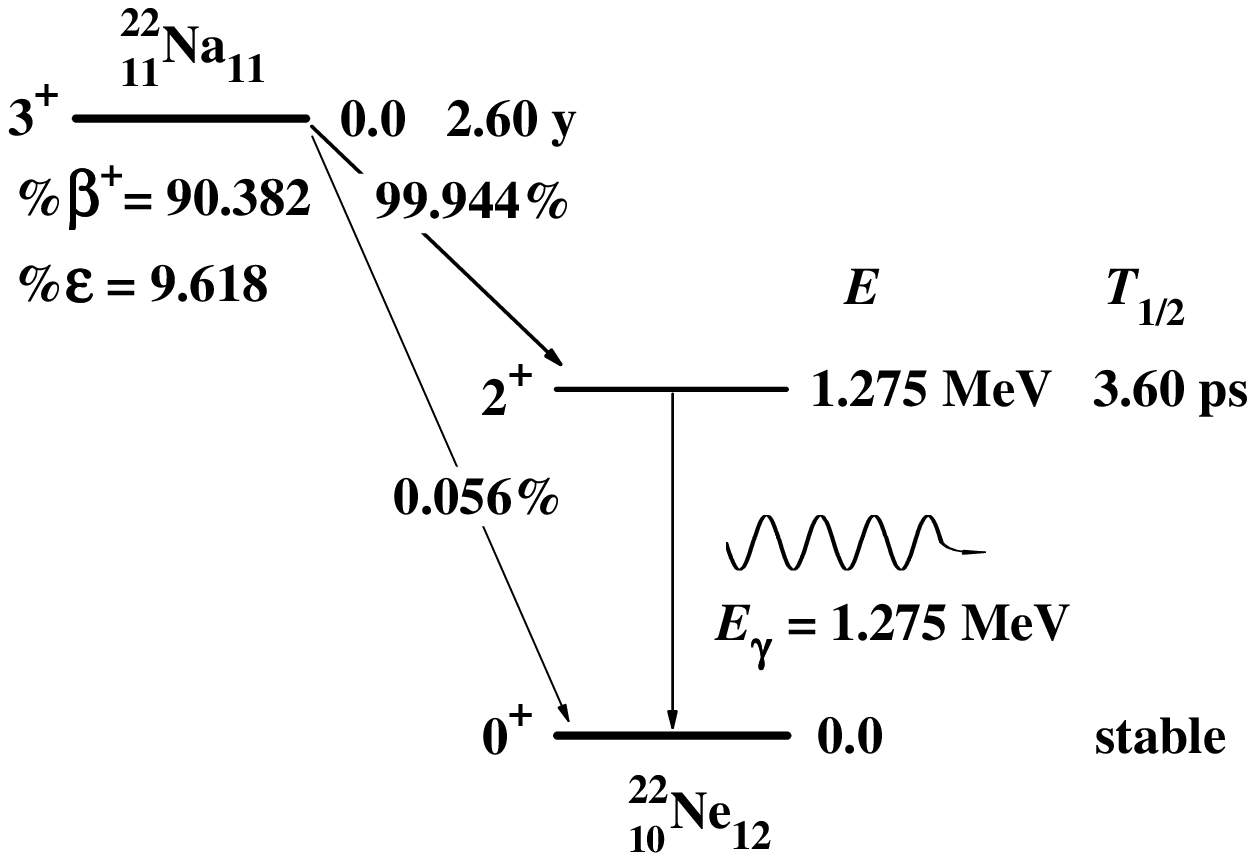} \\
      (a) & (b)
    \end{tabular}
    \caption{Schematic diagrams of the $^7$Be (a) and of the $^{22}$Na (b) spontaneous decay,
      the data are from \cite{Tilley2002} and \cite{Basunia2015}, respectively.}
\label{fig1}
\end{center}
\end{figure}

Both $^7$Be and $^{22}$Na are unstable radionuclides (see Fig.\ref{fig1}) and are well-known to physicists.
$^7$Be decays into stable $^7$Li via the electron capture mechanism with the half-life of 53~days. 
As the decay rate depends on the electronic density at the nucleus,
$^7$Be has been used in the studies of the effect of the chemical environment onto the nuclear processes
\cite{Segre1947,daudel1947,Tkalya2012}.
Besides, and even more importantly, $^7$Be is assumed to play a key role in the lithium yield of the big
bang nucleosynthesis for standard
cosmology via the neutron absorption reaction $^7$Be(n,p)$^7$Li, \cite{Cyburt2016,Damone2018}.
$^{22}$Na decays mainly by emitting positrons and the $\gamma$-radiation into the stable $^{22}$Ne isotope,
and have the half-life of 2.6 years. It is important for the nucleosynthesis problem \cite{Koehler1988Na22}
and also is used in the positron emission tomography (PET). 

In Table~\ref{tab1}, we summarize the neutron capture properties of the two radionuclides 
and compare them with the properties of $^{10}$B.
%In what follows, we will compare side-by-side physical grounds of BNCT and BeNCT,
%see Tables~\ref{tab1},\ref{tab2}. 
%
\begin{table}[t]
  \begin{center}
  \caption{Neutron capture properties \cite{Carlson2018,Tomandl2019,Koehler1988Na22}}
  \label{tab1}
  \begin{tabular}{c|c|c}
    \hline \hline
    \multicolumn{3}{c}{nuclide properties} \\
    \hline
    $^{10}$B$\left(J^\pi=3^+\right)$ stable   &
    $^{7}$Be$\left(\frac{3}{2}^-\right)$ ${53.2~\textrm{d}} \atop \longrightarrow $ $^{7}$Li &
    $^{22}$Na$\left(3^+\right)$ ${2.60~\textrm{y}} \atop \longrightarrow$ $^{22}$Ne\\
    \hline \hline
    \multicolumn{3}{c}{main reaction channels:}  \\
    \multicolumn{3}{c}{"0"--the ground state of daughter nucleus; "1" -- the first excited state}  \\
   \hline
   $^{10}$B+n $\rightarrow$   $^{11}$B$^*$ $\rightarrow$   & $^{7}$Be+n $\rightarrow$ $^{8}$Be$^*$ $\rightarrow$ &
                                                           $^{22}$Na+n $\rightarrow$ $^{23}$Na$^*$ $\rightarrow$ \\
   (93.3\%)$^7$Li$_1$+$\alpha_1$+$\gamma$      & (1.2\%)$^7$Li$_1$+p$_1$+$\gamma$ &
                                                (99.15\%)$^{22}$Ne$_1$+p$_1$+$\gamma$ \\
   (6.7\%) $^7$Li$_0$+$\alpha_0$               & (98.8\%) $^7$Li$_0$+p$_0$  &
                                                (0.85\%) $^{22}$Ne$_0$+p$_0$     \\
   \hline \hline
   \multicolumn{3}{c}{kinetic energy of heavy particle products in the predominant channel, [MeV]} \\
   \hline
   $0.84(^7$Li$_1)$ + $1.47(\alpha_1)$  &  $0.21(^7$Li$_0)$ + $1.44($p$_0)$ & $0.10(^{22}$Ne$_1)$ + $2.25($p$_1)$ \\
     \hline\hline
     \multicolumn{3}{c}{total neutron capture cross section}  \\
%     \hline
     \multicolumn{3}{c}{$\sigma_{\mathrm nX}\sqrt{E_\mathrm{n}}$, [kb eV$^{1/2}$] }  \\
   \hline
   0.6114 & 7.0 & 5.0 \\
   \hline   
   \multicolumn{3}{c}{$\sigma_{nX}\left(E_n[\textrm{eV}]\right)$ [kb] } \\
   \hline
 $\sigma_{\mathrm n\alpha}(0.0253) = 3.844$  & $\sigma_{\mathrm np}(0.0253) = 44$ & $\sigma_{\mathrm np}(0.0253) = 31$\\ 
 $\sigma_{\mathrm n\alpha}(1.0) = 0.611$     & $\sigma_{\mathrm np}(1.0) = 7.0$   & $\sigma_{\mathrm np}(1.0) = 5.0$ \\
 $\sigma_{\mathrm n\alpha}(10) = 0.193$      & $\sigma_{\mathrm np}(10) = 2.2$    & $\sigma_{\mathrm np}(10) = 1.6$\\
 $\sigma_{\mathrm n\alpha}(100) = 0.0611$    & $\sigma_{\mathrm np}(100) = 0.7$  & $\sigma_{\mathrm np}(100) = 0.5$ \\
   \hline\hline
  \end{tabular}
  \end{center}
\end{table}
The neutron capture reactions pass via an intermediate compound nucleus in an excited state,
-- $^{11}$B, $^8$Be, or $^{23}$Na, correspondingly. The compound nuclei then decay into the daughter nuclei
  ( the $^7$Li in the case of B and Be and the $^{22}$Ne for Na) by emitting a heavy particle 
-- an $\alpha$-particle for boron and a proton in the cases of beryllium and sodium.
With the probability of 93.3\%, the $^{11}$B$^*$ decays into the first excited state of $^7$Li
emitting the characteristic 478~keV $\gamma$-radiation (see, Fig.\ref{fig1}a).
Also $^{23}$Na$^*$ mainly (99.15\%) decays into the first excited state of its daughter nucleus, $^{22}$Ne,
emitting a 1.275~MeV photon (see, Fig.\ref{fig1}b).
In contrast, the $^8$Be$^*$ decays predominantly (98.8\%) into the ground state of $^7$Li.
The cross section for the $^{10}$B(n,$\alpha\gamma$)$^7$Li reaction is well determined
(see recent update in \cite{Carlson2018}). It follows Bethe's $\propto 1/v$ law with the accuracy better than 5\%
for the neutron incident energies $E_n<100$ keV, so that
$\sigma_{\mathrm n\alpha}\sqrt{E_{\mathrm n}} = 0.6114$~kb~eV$^{1/2}$.
The cross section of the $^{7}$Be(n,p)$^7$Li follows the $\propto 1/v$ law for a much smaller neutron
kinetic energy range, $E_{\mathrm n}<100$ eV, \cite{Damone2018}. Moreover, there is inconsistency (of more than 20\%)
between the values of $\sigma_{\mathrm n\alpha}$ obtained by different research groups. Thus, for the thermal neutrons
($E_n=0.0253$~eV) the neutron capture cross section varies from $38.4\pm0.8$~kb \cite{Koehler1988} to
$52.3\pm5.2$~kb \cite{Damone2018}. This discrepancy is still under debates \cite{Tomandl2019}.
We have adopted the latest result, \cite{Tomandl2019},
so that $\sigma_{\mathrm n\alpha}\sqrt{E_\mathrm{n}} = 7.0$~kb~eV$^{1/2}$ in the above mentioned neutron energy range.
The $^{22}$Na(n,p$\gamma$)$^{22}$Ne reaction also follows the $\propto 1/v$ law till the neutron energies,
$E_{\mathrm n}<100$ eV, so that  $\sigma_{\mathrm n\alpha}\sqrt{E_\mathrm{n}} = 5.0$~kb~eV$^{1/2}$ \cite{Koehler1988Na22}.

It is interesting to note that, besides $^{7}$Be and $^{22}$Na,
there are no other reasonably stable nuclides that have the capture cross section
for the thermal and the epithermal neutrons larger than that for $^{10}$B and emit a high LET radiation
suitable for NCT, \cite{Carlson2018}. There are very strong neutron absorbers like $^{135}$Xe and $^{113}$Cd,
but via the $(n,\gamma)$ reaction, i.e. emitting $\gamma$-radiation not suitable for the NCT. 

\section*{Neutron capture energy deposition}
For more direct comparison of the BNCT with potential BeNCT and NaNCT,
we estimate the energy deposition after neutron capture in water,
see Table~\ref{tab2}, considering only the predominant reaction channels.
\begin{table}[t]
  \begin{center}
    \caption{BNCT (left) vs BeNCT (center) vs NaNCT (right): the energy deposition in water.
      The values "per cell" stand for per mass, $M_\mathrm{cell}=2.3$~ng; 
      $E_\mathrm{n}$ denotes the neutron kinetic energy.}
  \label{tab2}
  \begin{tabular}{c|c|c}
    \hline \hline
    \multicolumn{3}{c}{high LET radiation energy, $E_\mathrm{t}$, MeV} \\
    \hline
    2.31  & 1.65 & 2.35 \\
    \hline \hline 
    \multicolumn{3}{c}{ranges [$\mu$m] of the high LET radiation} \\
    \hline
    $R_\alpha = 8$ , $R_\mathrm{Li}=5$ & $R_\mathrm{p} = 44$ , $R_\mathrm{Li}=2$ & $R_\mathrm{p} = 92$ , $R_\mathrm{Ne} < 1$ \\
    \hline \hline
    \multicolumn{3}{c}{number of reactions per cell ($N_\mathrm{r}$)} \\
    \multicolumn{3}{c}{required for the dose of 20 Gy} \\
    \hline
    120  & 170  & 120\\   
    \hline \hline
    \multicolumn{3}{c}{nuclide concentration, $c(E_\mathrm{n})$ [$\mu$g~g$^{-1}$]} \\
    \multicolumn{3}{c}{(in parenthesis -- the number of nuclides per cell, $N_\mathrm{cell}(E_\mathrm{n})$) } \\
    \multicolumn{3}{c}{required for the dose of 20 Gy absorbed} \\
    \multicolumn{3}{c}{during 1 hour exposition to neutron flux of 10$^{10}$~s$^{-1}$cm$^{-2}$} \\
    \hline
    \multicolumn{3}{c}{$E_\mathrm{n}=0.0253$ eV} \\
    6.5 (0.9$ \times 10^{9}$) & 0.56 (0.11$ \times 10^{9}$) & 1.7 (0.11$ \times 10^{9}$)\\
    \hline
    \multicolumn{3}{c}{$E_\mathrm{n}=1.0$ eV} \\
    41 (5.7$ \times 10^{9}$) & 3.5 (0.69$ \times 10^{9}$) & 11 (0.68$ \times 10^{9}$) \\
    \hline
    \multicolumn{3}{c}{$E_\mathrm{n}=10.0$ eV} \\
    130 (18$ \times 10^{9}$) & 11 (2.2$ \times 10^{9}$) &  34 (2.1$ \times 10^{9}$) \\
    \hline
    \multicolumn{3}{c}{$E_\mathrm{n}=100.0$ eV} \\
    410 ($57 \times 10^{9}$) & 35 ($6.9 \times 10^{9}$)  & 110 (6.8$ \times 10^{9}$) \\
    \hline
  \end{tabular}
  \end{center}
\end{table}
The $\gamma$-radiation emitting in the predominant channels for BNCT and NaNCT (Table~\ref{tab1}) 
is a low linear energy transfer (LET) radiation, i.e. it has low probability to produce ionization
in the vicinity of the decaying nuclide, rather escaping from the tumor region.
A part of the decay energy may be transformed into the electronic
excitation of the daughter nucleus via different mechanisms
(e.g., the internal conversion, the coulomb excitation by the emitting charged particles,
the shake-off due to the instant change of the nucleus charge and of its momentum)
and can be released in the forms of Auger electrons and photons.
But for the light elements considered here (Be, B and Na), the energy released in this form and
its biological effects may probably be neglected for the NCT, in contract with the heavy elements
like Gd, where the internal conversion is important \cite{Enger2013}.
Thus, only the kinetic energy, $E_\mathrm{t}$, transferred to charged heavy particles (high LET radiation)
is useful for the NCT and are reported in the top of Table~\ref{tab2}. 
%This energy, also is larger for $^{10}$B
%$($E_{t}=2.31$~MeV for the predominant channel)
%than for $^{7}$B ($E_t=1.65$~MeV for the predominant channel).
%
%
The ranges of the protons and of the $\alpha $-particles in water are taken from
\cite{Berger2017}, while those for the Li and Ne  ions are from \cite{Northcliffe1970}.

To relate the radiation doses in water to the biological context, we
will consider the typical cell linear dimension, $L_\mathrm{cell}=13$~$\mu$m,
and the typical cell mass, $M_\mathrm{cell}=2.3$~ng, \cite{supplmat}.
Also, we will assume $D_\mathrm{t}=20$~Gy in tumor tissues for a typical dose required for the NCT in
a single run \cite{Barth2005,Mitin2014}.
To compare the nuclides, we use the number of neutron capture reaction per cell, $N_\mathrm{r}$,
required to release locally the absorbed radiation dose $D_\mathrm{t}$,  \cite{supplmat}.
This quantity is independent on the neutron capture probability and reflects only the difference
(among the nuclides) in the total energy of high LET radiation following the neutron capture.
Only a part of the nuclides in the cell absorbs neutrons and contribute to the radiation dose,
depending on the neutron flux, $J_\mathrm{n}$, the neutron kinetic energy, $E_\mathrm{n}$,
the absorbtion cross section, $\sigma_\mathrm{n}(E_\mathrm{n})$ and the time of the irradiation.
Therefore, the required number, $N_\mathrm{cell}$, of nuclides per cell to give the absorbed dose
$D$ while exposing for time $t$ to neutron flux $J_\mathrm{n}$ is a more relevant quantity
to compare nuclides for the NCT. Alternatively to $N_\mathrm{cell}$,
the relative concentration, $c$, of the nuclide in the target
(in the units of ppm or $\mu$g of the nuclide per 1 g of tissue) is often used.
Both $N_\mathrm{cell}$ and $c$ are shown in Table~\ref{tab2} for $D_\mathrm{t}=20$~Gy, $t=1$h,
$J_\mathrm{n}=10^{10}$~s$^{-1}$cm$^{-2}$ and for different neutron kinetic energies.

\section*{BeNCT and NaNCT: advantage and perspective}
Based on the data and the estimates shown in the tables, we can discuss pros and cons of the potential
BeNCT and NaNCT in comparison to the existing BNCT.
First, one finds that the neutron absorption cross section
for $^7$Be is more than 10 times (see, Table~\ref{tab1}) larger and for $^{22}$Na is more than 8 times larger
than that for $^{10}$B for $E_\mathrm{n}<100$~eV.
This principal advantage of the $^7$Be is somewhat counterbalanced by a lower (by 30\%)
released kinetic energy of the high LET radiation (see, Table~\ref{tab2}). For $^{22}$Na
the kinetic energy release is almost the same as for  $^{10}$B. 
In the end, the required number of $^7$Be nuclides per cell, $N_{\rm cell}$ is the same as for  $^{22}$Na, and
it is at least 8 times smaller that that for the boron, at the same neutron flux and the exposition time.
For example, for thermal neutrons, one would need $0.9\times 10^9$ $^{10}$B
nuclei per 2.3 ng of water (typical cell) to obtain 20 Gy of the absorbed radiation dose after
1 hour of exposition to $J_\mathrm{n}=10^{10} $s$^{-1}$cm$^{-2}$ neutron flux,
while only $1.1\times 10^8$ of $^7$Be or of $^{22}$Na is required.
This ratio remains valid till the neutron kinetic energy of 100 eV.
As the requirement of sufficient concentration of the nuclide is one of the key issues in BNCT
\cite{Barth2018}, the reduction by the factor of 8 in case of $^7$Be and of $^{22}$Na may be crucial for the
success of the NCT. The larger cross section would also make it possible to bring the required dose
to the tumors more deeply buried in the body \cite{supplmat} than for BNCT,
thus increasing the potential applicability of the therapy.

It is worth to note that the main ionizing particles are different
for boron and for the new nuclides ($\alpha$-particles and protons, respectively),
so that the biological effect may be different at the same total kinetic energy.
On one side, the $\alpha$-particle has a shorter range and, if the $^{10}$B is located close
to the cell nucleus, it will release its kinetic energy within several microns, creating maximum radiation
damages directly to the DNA. The proton radiated by $^7$Be or $^{22}$Na after the neutron capture
travels tens of microns and can release its kinetic energy to several cells,
thus reducing the biological damages to each one.
On the other hand, if the activated nuclide is located in the cell far from the nucleus, the $\alpha$-particle
can stop before reaching the important organelles and produces only a small radiation damage. The proton, however,
has a higher probability to hit a cell nucleus while crossing several adjacent cells, and thus might be more efficient.
Which of the above contributions dominates in the biological effect of the NCT will also depends
on the mode the nuclide is accumulated inside the tumor cells. This is one key question for future
experimental studies.

Another way to profit from the larger neutron capture cross section for the new nuclides is to reduce the
intensity of neutron flux, again by a factor of 8, while keeping the nuclide concentration
in tumor the same as for $^{10}$B. This would give two novel potential benefits to the BeNCT and the NaNCT.
First, the probability of activation of the nuclei in the healthy tissues will be reduced by 8 times
compared to BNCT, thus reducing the adverse side effects.
This should have very serious implications, for example,
when treating the brain glioblastoma and other brain cancers where the tumor is embedded deeply
inside the brain tissue \cite{Busse2003,Barth2012}. One needs to apply  epithermal neutrons
to reach the tumor in the case. 
On the other hand, the healthy brain tissue is very sensitive to the radiation damage
coming from epithermal neutrons, so that any significant reduction of the necessary neutron flux would
be very beneficial for the cancer therapy of the brain tumors.
Second, a lower required intensity ($10^{8-9}$s$^{-1}$cm$^{-2}$) of the neutron beam would make it
possible to use a smaller scale and cheaper accelerator-based neutron sources \cite{Kiyanagi2018,Anderson2016}.
This would make the NCT economically more efficient and more widely used.

Positrons, from the spontaneous decay of $^{22}$Na can be used in the positron emission tomography (PET).  
But also the characteristic $\gamma$-radiation (478 keV) from spontaneous decay of $^7$Be  (see, Fig.~\ref{fig1}a)
and the $\gamma$ (1.28 MeV) from the spontaneous decay of $^{22}$Na (see, Fig.~\ref{fig1}b)
can be utilized very favorably in the standard single photon emission computer tomography (SPECT) \cite{Patton2008}
to trace the accumulation of the nuclide before the neutron irradiation. This is also an important advantage
over the BNCT, where the control of the nuclide accumulation is difficult. There, one labels additionally the
boron-containing compound with a radioactive nuclide, e.g., $^{18}$F, with subsequent positron emission tomography
(PET) \cite{Hanaoka2014}.

\section*{Open problems}
There are also new challenges on the way of $^7$Be and $^{22}$Na to the successful NCT.

$^{22}$Na is used for PET and it is already a commercial product.
On the 10-1000 ng scale necessary for experimental studies,  $^7$Be may be produced
as a byproduct on existing accelerator-based neutron sources utilizing, e.g., the proton induced spallation
reaction on $^{16}$O \cite{Maugeri2018}.
For potential industrial-scale application, a dedicated production infrastructure will be necessary.
In this respect, the idea (first suggested in \cite{Bayanov1998})
of an accelerator for NCT utilizing the reverse reaction, $7$Li(p,n)$^7$Be,
and an intense (10 mA and more) proton beam to produce neutrons looks especially attractive.
Here, the beryllium isotope made during neutron generation can be further utilized for BeNCT,
thus also reducing the cost of radioactive waste management. 

The radiotoxicity problem of these radioactive isotopes is another one.
That is, one has to understand and to minimize the adverse effects of
the spontaneous radiation on the patient, but also on his/her family, the personnel and the environment
during the drug production, administration, excretion, and the radioactive waste management.
To assess the effects of the spontaneous radiation of the $^7$Be and the $^{22}$Na nuclides, 
we compare them \cite{supplmat} with the iodine radioisotope, $^{131}$I,
which emits similar radiation ($\beta$ and $\gamma$), and it is
widely used in the therapy of thyroid cancer \cite{Rubino2003}.
For the radiation therapy, one uses the amount of  $^{131}$I equivalent to
the activity ranging from 0.2 to more than 50 GBq \cite{Rubino2003}. As 4.6~GBq corresponds to
1~$\mu$g of the nuclide, one can take 1~$\mu$g as a typical mass 
per procedure with  $^{131}$I. The same mass of $^7$Be in the target would be necessary
to produce 20~Gy per hour in 1~g of tumor in BeNCT (see, Table~\ref{tab2}).
The external dose rate at the distance of 1 m from the source, $\dot{D}_\mathrm{ext}$,
which characterizes the radiation risk for the environment and the personnel,
is 2.5 times smaller for $^7$Be (0.09 mGy/h) than for the typical 
therapeutic amount of $^{131}$I. For $^{22}$Na, the necessary mass of the nuclide is 3 $\mu$g per 1~g
of tumor, and $\dot{D}_\mathrm{ext}=0.21$ mGy/h is similar to that of 1~$\mu$g of $^{131}$I.
The $\beta$ component of the internal spontaneous dose is absorbed locally, within 2 mm from the source. 
It produces the therapeutic effect in the tumor and it is harmful when the radionuclide
is the healthy tissues. In the case of $^{131}$I, the dose rate is $\dot{D}_\mathrm{int}^\beta = 360$~mGy/h
for 1~$\mu$g of the nuclide; it is 8 times lower for 3~$\mu$g of $^{22}$Na and it is absent for $^7$Be.
However, the main therapeutic effect of $^{22}$Na and of $^7$Be comes from the neutron capture reaction
and has the absorbed dose rate of 20 Gy/h, i.e., it is more than 50 time stronger than for $^{131}$I.
The $\gamma$ component of the internal dose is absorbed (by definition) within the sphere of $R=10$~cm
from the decaying nuclides and, most probably, outside the tumor. Therefore, it makes the radiation
damage to healthy tissues with the same dose rate of about $\dot{D}_\mathrm{int}^\gamma = 60$~mGy/h
per 1~$\mu$g of $^{131}$I or per 3~$\mu$g of $^{22}$Na; the effect is 2.5 time smaller for
1~$\mu$g of $^7$Be.
Therefore, we conclude that if administered in the same amount as $^{131}$I
for the thyroid cancer therapy (i.e., in the range of micrograms),
the radiation risk of from spontaneous decay of $^7$Be and $^{22}$Na would be at the same level or smaller
then that for $^{131}$I. Thus, the nuclides can be managed and administered by following standard security
procedure and technologies for radiopharmacuticals. 
But simultaneously the therapeutic effect for the tumor can be much stronger and with better temporal
control for NCT than for the standard radionuclide treatment.

However, the problem, and this is the principal challenged for the future NCT research,
is how to provide the targeting, the accumulation and the pharmacokinetic timescales for $^7$Be
and $^{22}$Na at least at the same level as in the case of $^{131}$I therapy for thyroid cancer.
In other words, one needs a pharmaceutical which binds the radionuclide, transport it selectively into
a specific cancerous tissue providing the nuclide concentradion of 1-3 ppm for several hours,
and then is excreted together with the remained non-activated nuclides within several days.
Of the total administered amount of several (maximum, few tens) micrograms of the nuclide per procedure
(the amount is limited by the spontaneous radiation risk), a substantial part (probably, more than 10\%)
should pass through the tumor. The above mentioned limitations of the total amount of nuclide
and the required pharmacokinetics would also eliminate the problem of chemical toxicity of the pure
elements, in particular, that for beryllium \cite{Betoxicity}. 
Besides, there is the second criterion of the nuclide accumulation selectivity
coming from the requirement to minimize the radiation risk under neutron irradiation -- 
the ratio of the nuclide concentration in the tumor and in the surrounding healthy
tissues must be more than 3 \cite{Moss2014}.

The earlier approach employed in BNCT consists of binding the active nuclide into
a soluble enough and low-toxic compound administered in large quantities to achieve
the desirable concentration in the target tissue. Two examples of this approach are
boronophenylalanine (BPA) and sodium borocaptate (BSH) \cite{Barth2018}.
However, the selectivity of accumulation for these compounds is not sufficient for the radionuclides.
For example, one had to administer around 10 mg of boron per 1 kg of patient's weight
\cite{Moss2014} to achieve the necessary concentration of the nuclide in tumor. Therefore, only a small
part of $^{10}$B ends up in tumor, while almost 1 g of that goes into healthy tissues and should be excreted.
Obviously, this approach does not fit the BeNCT and NaNCT due to the spontaneous radioactivity.
Another method of delivery can be based on recent developments in the nanomedicine \cite{Tran2017,Barth2018}
and  could be utilized for the new NCT.
This modular approach can comprise two or three of the following steps.
One first binds the nuclide in a stable compound, then encapsulates it into a
nanoparticle having sufficient solubility, pharmacokinetics and low toxicity,
and finally functionalizes the nanoparticles to target specific cells.
The nanomedicine approach can potentially provide a very high selectivity, and
it is especially suitable for NCT.
Here, there is no need to release the drug in the target tissues as in the case of chemically acting compounds
-- the ionizing radiation will exit the nanocage anyway. Moreover, it would be an advantage to keep
the non-activated nuclides tightly encapsulated during the whole procedure to eliminate the radiotoxicity
risk and control the excretion. A possible mean to bind beryllium or sodium in order to protect
it from the interaction with aqueous medium is to encapsulate it into fullerens (or sealed nanotubes).
We have shown recently \cite{Tkalya2012,Bibikov2013}, that
the energetic barrier for beryllium to cross the wall of a fulleren is 1-2 eV. Then, a poorly soluble fulleren
can be functionalized or encapsulated (into, e.g. a liposome) to improve the pharmacokinetics,
and be further functionalized to target the specific tissue,
by following existing nanomedicine techniques \cite{Tran2017,Barth2018}.
Otherwise, recent developments in the beryllium-organic chemistry \cite{Perera2017,Iversen2015} could
provide other options for coordination of beryllium in organic chelators.
We are not intended to discuss in details the chemistry issues in BeNCT and NaNCT here.

\section*{Conclusion}
Instead, we have put forward physical arguments in favor of potential use of $^7$Be and  $^{22}$Na
for the neutron capture therapy. We have discussed its advantages vs existing BNCT and
also the limitations and new challenges arising from the radioactivity and the production issues.
The use of new nuclides alone or in combination with traditional $^{10}$B and/or $^{157}$Gd would
open up a new way in the neutron capture therapy, unexplored before.
There no other reasonably stable nuclides that have the capture cross section
for the thermal end epithermal neutrons larger than that for $^{10}$B and emit a high LET radiation
suitable for NCT.
By combining BeNCT and NaNCT with SPECT (or NaNCT with PET) one would obtain a real theranostics
-- simultaneous diagnostics and targeted therapy of cancerous tissues,
-- an important ingredient of modern personalized medicine.

\begin{acknowledgments}
{\bf Funding:} EVT was supported by a grant of the Russian Science Foundation (Project No 19-72-30014).
{\bf Author contributions:} VIK put forward the original idea of using $^7$Be. IVB wrote the initial draft.
AVB, IVB, EVT and MC performed calculations, data collection and analysis.
All the authors reviewed and edited the final version of the manuscript.
\end{acknowledgments}
         
\bibliography{benct}

%Here you should list the contents of your Supplementary Materials -- below is an example. 
%You should include a list of Supplementary figures, Tables, and any references that appear only in the SM. 
%Note that the reference numbering continues from the main text to the SM.
% In the example below, Refs. 4-10 were cited only in the SM.     
\section*{Supplementary materials}
\subsection*{Materials and Methods}
\paragraph{Typical cell size and mass.}  
The average volumes of human cells range from 30 to $4\times 10^6$~$~\mu$m$^3$,
depending on the tissue \cite{bionumbers}; the corresponding linear dimensions range from 3 to 160 $\mu$m.
For the estimates, we will assume the typical cancer cell volume, $V_\mathrm{cell}=2300$~$\mu$m$^3$,
and the typical cell linear dimension, $L_\mathrm{cell}=13$~$\mu$m, which corresponds to HeLa cells often used in cancer
{\it in vitro} studies \cite{bionumbers}. Then, we assume the mass density of the cell,
$\rho_\mathrm{cell} = 1$~g~cm$^{-3}$, so that the typical cancer cell's mass, $M_\mathrm{cell}=2.3$~ng.

\paragraph{Estimation of the absorbed radiation dose under the neutron irradiation.}
The dose of hight LET radiation absorbed in a cell reads, $D = E_\mathrm{t} N_\mathrm{r} /  M_\mathrm{cell}$, where
$N_\mathrm{r}$ is the number of the neutron capture reactions happened. Thus, the required $N_\mathrm{r}$
to release locally in the cell the absorbed radiation dose $D$ is, $N_\mathrm{r}=D M_\mathrm{cell} / E_\mathrm{t}$.
Here, we underline again, we have assumed that the gamma-radiation after neutron
capture has a small probability to interact and ionize matter within the tumor region ($< 10$~cm), so that the
locally deposed energy is only that of the charged particles, $E_t$. The corresponding estimates for
the BNCT, the BeNCT and the NaNCT are shown in Table~\ref{tab2},
assuming $D=20$~Gy in tumor tissues for a typical dose required for the radiation
therapy in a single run \cite{Barth2005,Mitin2014}.

The absorbed dose rate in a cell under a constant neutron flux $J_\mathrm{n}$ reads,
$\dot{D}(t)=E_\mathrm{t} \dot{N}(t) / M_\mathrm{cell}$, where
$\dot{N}(t)=N_0\exp\left(-t/ \tau_\mathrm{n}\right)/ \tau_\mathrm{n}$
is the number of the neutron capture reactions in the cell per second,
$N_0$ is the initial (at $t=0$) number of nuclide particles per cell;
$\tau_\mathrm{n}=1/ (\sigma_\mathrm{n} J_\mathrm{n})$ is the
reaction time constant; $\sigma_\mathrm{n}$ is the neutron capture cross section.
For typical values of $\sigma_\mathrm{n}$ (see, Table~\ref{tab1}) and of the  neutron flux,
$J_\mathrm{n} < 10^{13} $s$^{-1}$cm$^{-2}$, the reaction time constant ($\tau_\mathrm{n} > 10^6$~s)
is much larger than the reasonable
duration of the NCT procedure (up to few hours). Therefore, if $N_0$ is not changing significantly
during the neutron irradiation
procedure due to the accumulation and the excretion, the dose rate is time independent,
and the absorbed dose for time $t$ of the neutron exposure reads,
$D(t)= t E_\mathrm{t} N_0 \sigma_\mathrm{n} J_\mathrm{n} / M_\mathrm{cell}$.
Then, the number of nuclides per cell to have the absorbed dose $D$ while exposing for time $t$ to neutron
flux $J_\mathrm{n}$, reads, $N_\mathrm{cell}=N_\mathrm{r}/(t\sigma_\mathrm{n}J_\mathrm{n})$.
The corresponding required mass concentration of the nuclide,
$c\equiv M(\mathrm{nuclide})/M_\mathrm{cell}$, reads,
$c=D \mu /(E_\mathrm{t} N_\mathrm{A} \sigma_\mathrm{n}J_\mathrm{n}t)$, where $\mu$ is the molar
mass of the nuclide and $N_\mathrm{A}$ is the Avogadro constant. 
In Table~\ref{tab2}, we show the estimated concentration of nuclide (both $N_\mathrm{cell}$ and $c$)
required to obtain the absorbed radiation dose of $D=20$~Gy in water after 1~hour
exposition to $J_\mathrm{n}=10^{10}~$s$^{-1}$cm$^{-2}$ neutron flux for thermal and epithermal neutron energies.

Once the neutrons enter the target (a water phantom or a human body) with initial energy $E_\mathrm{n}^0$,
they start to scatter on the atoms (mainly elastically and on hydrogens) and to slow down.
The average neutron energy in the target, $E_\mathrm{n}(x)$,
decreases with the depth $x$ from the initial energy $E_\mathrm{n}(0)$  at $x=0$ till
the thermal energy $3/2 k_\mathrm{B}T$ at a given temperature $T$,
and the neutron capture cross section, $\sigma_\mathrm{n}(E_\mathrm{n}(x))$, for the active nuclide increase with $x$
reaching the value for the thermal neutrons.
Besides, neutrons change their direction after each collision, so that the initially collimated beam will disperse
with the distance traveled in the target. Therefore, the neutron flux $J_\mathrm{n}(x)$ along the initial beam direction
decreases  with the depth $x$.  As the absorbed dose is proportional to the product,
$D(x) \propto \sigma_\mathrm{n}(E_\mathrm{n}(x)) J_\mathrm{n}(x)$, it also depends on the depth $x$ even at the
uniform concentration of the active nuclide. If the incident neutrons are already thermalized,
their average energy and the capture cross section are constant, so that the dose decreases with the depth due to the
decrease of the average flux. If the initial neutrons have a higher energy (0.5~eV-10~keV), the absorbed dose, $D(x)$,
first increases with $x$ together with the neutron capture cross section,
then reaches its maximum at a certain distance $x_0$ (1-4 cm) \cite{Kiyanagi2018}, and then decreases due to the
flux angular dispersion. The interval of the depth, $[x_\mathrm{min}, x_\mathrm{max}]$ around $x_0$, where the absorbed
dose is higher than a certain critical value (e.g., 20 Gy/h) at a given nuclide concentration is suitable for the NCT.
For the $^7$Be and the $^{22}$Na nuclides, having the neutron capture cross section larger than that for $^{10}$B
by a factor of 10, the maximum working depth, $x_\mathrm{max}$, can be much larger than $x_0$ and can reach 10-20 cm
required for potential therapy of deep sitting tumors.

\paragraph{The absorbed dose from the spontaneous radiation.}
To estimate the effects of the spontaneous radiation of the $^7$Be and the $^{22}$Na nuclides, 
both for the patient and for the personnel, we compare them
with the iodine radioisotope, $^{131}$I, which emits similar radiation and also is
widely used in the therapy of thyroid cancer \cite{Rubino2003}, see Table~\ref{tabs1}.
\begin{table}[t]
  \begin{center}
    \caption{Comparison of the radiation effects of the spontaneous decay}
  \label{tabs1}
  \begin{tabular}{c|c|c}
    \hline \hline
    \multicolumn{3}{c}{Nuclide (half-life)} \\
    \hline
    $^{131}$I (8.0~d) & $^7$Be (53~d)& $^{22}$Na (2.60~y) \\
    \hline \hline
    \multicolumn{3}{c}{Mass-specific activity, $A_\mathrm{m}$ [GBq/$\mu$g]}\\
    \multicolumn{3}{c}{radiation type, its energy and the abundance } \\
    \hline
    4.6  & 13 & 0.23 \\
    $\gamma$, 370 keV - average  & $\gamma$, 478 keV, f=0.10 & $\gamma$, 1280 keV, f=1 \\
    $\beta^-$, 570 keV - average &                           & $\beta^+$, 540 keV , f=0.9 \\
                                 &                           & $\gamma$, 511 keV , f=1.8 \\    
    \hline \hline 
    \multicolumn{3}{c}{Gamma constant at 1~m distance, $\Gamma$ [mGy/h/GBq]} \\
    \hline
    0.05  & 0.007 & 0.30 \\
    \hline \hline
    \multicolumn{3}{c}{External absorbed dose rate, $\dot{D}_\mathrm{ext}$, [mGy/h]} \\ 
    \multicolumn{3}{c}{per 1 $\mu$g of nuclide at 1 m distance } \\
    \hline
    0.23 & 0.09 & 0.07 \\   
    \hline \hline
    \multicolumn{3}{c}{Internal absorbed dose rate per 1 $\mu$g of nuclide, $\dot{D}_\mathrm{int}$, [mGy/h]} \\
    \multicolumn{3}{c}{contribution for each radiation type is specified separately } \\  
    \hline
    60 - $\gamma$    &  25 - $\gamma$ & 18 - $\gamma$ \\
    360 - $\beta$    &                & 15 - $\beta$  \\
    Total: 420       &                & Total: 33 \\
    \hline
  \end{tabular}
  \end{center}
\end{table}
There are several standard physical characteristics of the spontaneous decay.
The iodine isotope has the shortest lifetime, $T_{1/2}$, of 8 days, while the sodium isotope's half-life is
the longest among the three nuclides and equals 2.6 years. The activity of a radionuclide
(the number of decays per second) of initial mass $M$, after time $t$ reads,
\begin{eqnarray}
   A(t) & = & A_\mathrm{m} M \exp\left(-\ln(2)\frac{t}{T_{1/2}}\right), \nonumber
\end{eqnarray}
where $A_\mathrm{m}=\ln(2)N_\mathrm{A}/(\mu T_{1/2})$ is the nuclide-mass-specific activity,
$\mu$ is the molar mass. Table~\ref{tabs1} shows that the mass-specific activity for the  $^7$Be is
the largest despite the half-life for $^{131}$I is shorter; apparently, it is due to the almost 20 times larger
isotope mass for the iodine.

To characterize the external (personnel and environmental)
radiation risk of a radionuclide, one ordinary calculates the  absorbed radiation dose
rate at a certain distance $d$ from a point-like source of activity $A$.
The distance is typically large enough (e.g., 1 m) so that all the $\beta$ radiation
is absorbed well before $d$, and only $\gamma$-radiation contributes to the ionization.
The external $\gamma$-radiation dose rate reads,
\begin{eqnarray}
  \frac{d D_\mathrm{ext}}{d t} & = \Gamma \frac{A}{d^2}P_\mathrm{s}. \nonumber
\end{eqnarray}
Here, the gamma constant,
\begin{eqnarray}
  \Gamma & = & \frac{1}{4\pi} \frac{\mu_\mathrm{en}}{\rho} \sum_i E_\gamma^i f^i , \nonumber
\end{eqnarray}
is a standard characteristics of the external radiation risk of the nuclide at given activity.
The sum in the latter equation runs over the energies of all the emitted gammas weighted by the
corresponding abundance ($f^i$ is the average number photons of type $i$ per decay of the nuclide).
The energy attenuation coefficient, $\mu_\mathrm{en}$, determines the average relative energy loss
of the photon beam after passing distance $x$ in the material $E/E_0=\exp(-\mu_\mathrm{en} x)$.
As $\mu_\mathrm{en}$ is approximately proportional to the mass density, $\rho$, of the material,
one often introduces the mass-energy attenuation coefficient, $\mu_\mathrm{en}/\rho$.
In particular, for both air and liquid water, $\mu_\mathrm{en}/\rho \approx 0.03$~cm$^2$g$^{-1}$ with the
20\% accuracy for the photon energies from 60 keV to 2 MeV, \cite{nist_gamma}.
The gamma shielding factor, $P_\mathrm{s}$, describes reduction of the average energy of the emitted photons
due to the absorption and the inelastic scattering on the way from the source to the observation
point at the distance $d$. If the energy attenuation coefficient of the shielding material,
as it was discussed above in the text,
$\mu_{en}^\mathrm{s}$, weakly depends on the photon energy, the shielding factor can be estimated as follows,
$P_\mathrm{s}=\exp\left(-\mu_{en}^\mathrm{s}L\right)$, where $L \leq d$ is thickness of the shield.
In  air, $\rho_\mathrm{s}=0.001225$~ g~cm$^{-3}$, the characteristic energy attenuation length,
$\lambda=1/\mu_{en}^\mathrm{s}$ is about 270~m, so that at the distance below d=1 m, the shielding factor,
$P_\mathrm{s} = \exp(-d/\lambda) \approx 1$. 
The energies and the corresponding abundances of the radiation are shown in Table~\ref{tabs1}.
For  $^{131}$I, more than 10 different photons may be emitted \cite{unger1982specific},
we show only the average (weighted) value, $\sum_i E_\gamma^i f^i$, in the table.
The calculated $\Gamma$ constants reported in Table~\ref{tabs1} agree with the
values recommended in \cite{unger1982specific} within 20\% accuracy sufficient for the
nuclide comparison presented here. This also justifies the approximations accepted for
the simplified estimates of the absorbed doses.
The calculated external dose rate at 1 m per 1 $\mu$g of nuclide, $\dot{D}_\mathrm{ext}$, is reported
in Table~\ref{tabs1}. 

To quantify the possible effect of the spontaneous radiation on the patient, we calculate the average
absorbed dose rate within a sphere of water of radius $R=10$~cm around a point-like source.
The $\beta$-radiation from the spontaneous decay is completely absorbed within less than 2 mm from the source
\cite{Berger2017} releasing the energy, $\sum_i E_\beta^i f^i_\beta$. The photons on average loose only a part of their
initial energy equal to $(1-\exp(-\mu_\mathrm{en}R))\sum_i E_\gamma^i f^i$.
The internal absorbed dose rate is the sum of the two contributions,
$\dot{D}_\mathrm{int}=\dot{D}_\mathrm{int}^\gamma+\dot{D}_\mathrm{int}^\beta$, 
\begin{eqnarray}
  \frac{d D_\mathrm{int}^\gamma}{d t} & = & A
  \left(1-\exp\left(-\mu_\mathrm{en}R\right)\right)\sum_i E_\gamma^i f^i \frac{3}{4\pi R^3\rho} .  \nonumber \\
  \frac{d D_\mathrm{int}^\beta}{d t} & = & A \sum_i E_\beta^i f^i_\beta \frac{3}{4\pi R^3\rho} .  \nonumber
\end{eqnarray}
The calculated external dose rate per 1 $\mu$g of nuclide, $\dot{D}_\mathrm{int}$, is reported
in Table~\ref{tabs1} for the photons and the $\beta$-particles, separately.

The half-life, the mass-specific activity and the $\Gamma$ constant reflects different physical properties
of radioactive nuclides, and, separately, are not sufficient to characterize 
the risk of the spontaneous radiation. The mass-specific internal, $\dot{D}_\mathrm{int}$,
and external, $\dot{D}_\mathrm{ext}$, dose rates defined here are more suitable to compare radionuclides.

For the radiation therapy of thyroid cancer one uses the amount of  $^{131}$I equivalent to
the activity ranging from 0.2 to more than 50 GBq \cite{Rubino2003}. As 4.6~GBq corresponds to
1~$\mu$g (see Table~\ref{tabs1}) of the nuclide, we can take 1~$\mu$g as a typical mass 
per procedure with  $^{131}$I. The same mass of $^7$Be in the target would be necessary
to produce 20~Gy per hour in 1~g of tumor in BeNCT (see, Table~\ref{tab2}).
The external dose rate at 1 m, $\dot{D}_\mathrm{ext}$, characterizing the radiation risk for
the environment and the personnel, is 2.5 time smaller for $^7$Be (0.09 mGy/h) than for the typical
therapeutic amount of $^{131}$I. For $^{22}$Na, the necessary mass of the nuclide is 3 $\mu$g per 1~g
of tumor, and $\dot{D}_\mathrm{ext}=0.21$ mGy/h is similar to that of 1~$\mu$g of $^{131}$I.

The $\beta$ component of the internal spontaneous dose is absorbed locally. 
It produces the therapeutic effect in the tumor and it is harmful when the radionuclide
is the healthy tissues. In the case of $^{131}$I, the dose rate is $\dot{D}_\mathrm{int}^\beta = 360$~mGy/h
for 1~$\mu$g of the nuclide. It is 8 times lower for 3~$\mu$g of $^{22}$Na and it is absent for $^7$Be.
However, the therapeutic effect of $^{22}$Na and $^7$Be comes from the neutron capture reaction
and has the absorbed dose rate of 20 Gy/h, i.e., it is more than 50 time stronger than for $^{131}$I.

The $\gamma$ component of the internal dose is absorbed (by definition) within the sphere of $R=10$~cm
from the decaying nuclides, and most probably, outside the tumor. Therefore, it makes the radiation
damage to healthy tissues with the same dose rate of about $\dot{D}_\mathrm{int}^\gamma = 60$~mGy/h
per 1~$\mu$g of $^{131}$I or per 3~$\mu$g of $^{22}$Na; the effect is 2.5 time smaller for
1~$\mu$g of $^7$Be.

\end{document}